\documentclass[a4paper]{article}

\usepackage{INTERSPEECH2021}
\usepackage{float}
\usepackage{microtype}
\usepackage{multirow}
\usepackage{hhline}
\usepackage{adjustbox}
\usepackage{subcaption}

\title{Generalized Spoofing Detection Inspired from Audio Generation Artifacts}
\name{Yang Gao$^{1,\star}$ \thanks{$\star$ Work performed during internship at AI Foundation}, Tyler Vuong$^1$,  Mahsa Elyasi$^2$, Gaurav Bharaj$^2$, Rita Singh$^1$}
\address{
  $^1$Carnegie Mellon University\;\;\;\;\;\;$^2$AI Foundation, USA}
\email{\{yanggao,tvuong,rsingh\}@andrew.cmu.edu \{gaurav,masha\}@aifoundation.com}

\begin{document}

\maketitle

\begin{abstract}

State-of-the-art methods for audio generation suffer from fingerprint artifacts and repeated inconsistencies across temporal and spectral domains. 
Such artifacts could be well captured by the frequency domain analysis over the spectrogram. Thus, we propose a novel use of long-range spectro-temporal modulation feature -- 2D DCT over log-Mel spectrogram for the audio deepfake detection. We show that this feature works better than log-Mel spectrogram, CQCC, MFCC, as a suitable candidate to capture such artifacts. We employ spectrum augmentation and feature normalization to decrease overfitting and bridge the gap between training and test dataset along with this novel feature introduction.  We developed a CNN-based baseline that achieved a 0.0849 t-DCF and outperformed the previously top single systems reported in the ASVspoof 2019 challenge. Finally, by combining our baseline with our proposed 2D DCT spectro-temporal feature, we decrease the t-DCF score down by 14\% to 0.0737, making it a state-of-the-art system for spoofing detection. Furthermore, we evaluate our model using two external datasets, showing the proposed feature's generalization ability. We also provide analysis and ablation studies for our proposed feature and results.

\end{abstract}
\noindent\textbf{Index Terms}: ASVspoof challenge, spoofing detection, 2D-DCT, modulation feature

\section{Introduction}


Audio deepfakes use deep learning and machine learning algorithms to generate or manipulate audio content with an intent to deceive. Such audio deepfakes are especially dangerous due to their innate embedding of biometrics, used in speech-based identity verification systems. State-of-the-art audio deepfake methods rely on voice conversion, text-to-speech synthesis, generative models, and neural vocoders \cite{wang2017tacotron, skerry2018prosodytransfer, gao2020interactiveTTS, gao2018voiceGAN, kaneko2018cycleganVC}. With these advances, the quality of deepfakes has significantly improved, making them a pernicious means to commit a wide variety of fraudulent activities --  identity theft and misinformation spread by untrained bad actors. Such techniques even outperform professional human impersonators and threaten automatic speaker verification (ASV) systems~\cite{gao2020detection}.


For better spoof attack detection in ASV systems, ASV spoof challenges \cite{wu2015asvspoof, wu2017asvspoof, kinnunen2017asvspoof, todisco2019asvspoof, nautsch2021asvspoof} have been created. In such challenges, the logical access (LA) consists of synthetically spoofed audio, which uses conventional signal processing and generative techniques that \cite{yu2017dnnfeatures, balamurali2019featurecomparison, kamble2020asvspooffeatures} propose the use of feature selection (e.g., Constant Q cepstral coefficients \cite{todisco2017cqcc}, MFCC, log-Mel spectrogram, etc.) to search for the best features for spoof detection. However, these features have been developed for generic tasks, such as automatic speech recognition (ASR) and sound-based event detection, etc. They may not capture the fundamental differences between real and fake speech well. Further, the choice of feature selection can be influenced by audio datasets and is inconsistent. For better generalization, as noted in \cite{gao2020detection}, unlike real speech, machine-generated speech consists of signature artifacts that can be leveraged for spoof detection. They propose a lightweight model with several human speech characteristics features and achieve comparably higher accuracy.


In computer vision, generative adversarial networks (GANs) \cite{goodfellow2014GAN} are a popular choice for image generation. Such methods have associated ``fingerprint''~\cite{yu2019GANfingerprints} and signal-domain~\cite{frank2020FrequencyAnalysisleveraging} artifacts that can be leveraged for detection and attribution studies. In speech synthesis, generative methods are used for feature learning from input linguistic features, while neural vocoders convert generated features into waveform outputs. Here, the audio is usually synthesized in frames or blocks of frames and has no cross-frame temporal consistency. This can lead to temporal modulation artifacts. Additionally, such methods are typically trained with element-wise mean-square-error losses in the Mel-Spectrogram domain \cite{gao2018voiceGAN, vuong2021modulation} and do not account for cross-frame consistency. Furthermore, speech is mainly encoded in the frequency ranges 0-4 kHz of auditory perception (based on the learning principles). There are associated artifacts with the generated outputs \cite{pons2021upsampling}, especially at high frequencies \cite{tian2016spoofinghl}.

Based on these observations for feature artifacts, we propose using long-range frequency analysis on log-Mel Spectrogram (in feature domain) for spoof detection. Since 2D-DCT features capture repeated patterns/artifacts by analyzing the joint spectro-temporal modulation frequencies, we introduce the novel use of global 2D-DCT on log-Mel Spectrograms, a long-range spectro-temporal feature, to capture audio deepfake artifacts. The spoof detection convolutional neural network (CNN) classifier that operates on log-Mel Spectrum consists of the features with limited receptive fields and focuses on finding local short/medium time patterns/correlations in the input audio. The proposed global 2D-DCT feature essentially forces the CNN classifier to learn from the input audio's long-term/global modulation patterns. These 2D-DCT features correspond to the long-term spectro-temporal modulations rather than localized ones. Therefore, we call this proposed feature \textit{global modulation} (Global M) feature. We show that the proposed feature detects deepfakes at a higher accuracy compared with the standard log-Mel features and could compensate our strongest baseline model to improve the overall detection performance further.

To summarize, in this paper, we compare the proposed global modulation features with traditional features such as MFCC, log-Mel, and CQCC and present the following novel contributions:
\begin{enumerate}
\item We propose a novel long-range spectro-temporal feature -- global modulation feature, for audio deepfake detection.
\item We further implement SpecAugment \cite{park2019specaugment} and feature normalization to reduce over-fitting and bridge the gap between training and test dataset from unseen attacks.
\item The resulting baseline system achieves the best tandem detection cost function (t-DCF) scores as single systems according to \cite{todisco2019asvspoof}. Furthermore, our proposed feature can compensate for this strong baseline to bring the t-DCF and the equal error rate (EER) down and achieve state-of-the-art performance on the ASVspoof challenge 2019 logical access (LA).
\end{enumerate}
Finally, the proposed global modulation feature also achieves a higher accuracy on general tasks, such as speaker verification, shown in Section 4.3.

\section{Related works}

\subsection{Audio deepfake detection}


The ASVspoof challenges \cite{wu2015asvspoof,kinnunen2017asvspoof,nautsch2021asvspoof} have raised efforts in fake speech spoofing attack countermeasures on ASV systems. Previous studies on anti-spoofing attacks on ASV systems and synthetic speech detection evaluate various features \cite{ kamble2020asvspooffeatures, sahidullah2015comparison} and deep learning models \cite{alzantot2019deepASVspoofpipeline} for detection performance. However, with the fast evolution of deepfake techniques, developing a detection system that is not constrained by the training data and can accurately detect new spoofed data generated from different or unseen deepfake algorithms is still challenging. 

In the ASVspoof challenge 2019 dataset, the logical access (LA) contains fake audio generated by multiple methods as in Table \ref{tab: ID types}. As reported in \cite{todisco2019asvspoof}, the best single system for LA data achieves a t-DCF metric \cite{todisco2019asvspoof} score of about 0.13 and an EER score of 5\%. The top-3 primary system (a weighted voting of multiple systems) achieves a t-DCF score of less than 0.1 and an EER of smaller than 3\%. 

There are also datasets for audio deepfake detection like FoR dataset \cite{reimao2019FoRdataset} and RTVCspoof dataset created using neural generation models as in \cite{subramani2020AIFlearning}. In our work, we also use these external datasets effectively as unseen test attacks to our proposed detection system. 


\subsection{Modulation features}



The modulation feature captures the longer time patterns in the signal, which are often ignored in MSE-based generation \cite{vuong2020learnableVT, vuong2021modulation}. Not only inspired by the generation artifacts, moreover, but the proposed feature is also global modulation feature that analyzes the joint long-range spectro-temporal modulation information. 

In \cite{chi2005multiresolutionspectrotemporal}, the importance of spectral and temporal modulation content of the auditory spectrogram is discussed. Here, filter banks selecting different spectro-temporal modulation parameters range from slow to fast rates temporally and from narrow to broad scales spectrally. The spectro-temporal receptive fields (STRFs) of these filters are related to human perception's auditory system. We also note that, from a physiological point of view, neurons in the primary auditory cortex of mammals are explicitly tuned to spectro-temporal patterns, e.g., spectro-temporal features, \cite{schadler2012spectro-asr}. Suthokumar et al.~\cite{suthokumar2018modulationreplay} analyze the temporal modulation by performing FFT analysis in each sub-band, and show the effectiveness of temporal dynamics for replay spoofing detection.

However, in previous studies, 2D-DCT was only used to calculate \textbf{local} spectro-temporal modulation, such as for robust automatic speech recognition (ASR) \cite{meyer2011asrmodulationcomparing}. Medium range modulation features were discussed in \cite{hermansky1994rasta, hermansky1999temporal} and long-range modulation was proposed in \cite{wu2013syntheticmodulation} -- but both only for the temporal domain. 
Our \textbf{global} modulation feature combines spectral (as MFCC) and temporal modulation information for better long-range feature modeling. To the best of our knowledge, such a long-range feature modeling has not been carried out in previous studies in speech.


\section{Experiments}

\subsection{Baseline model}

\begin{figure}[t]
  \centering
  \includegraphics[width=1\linewidth]{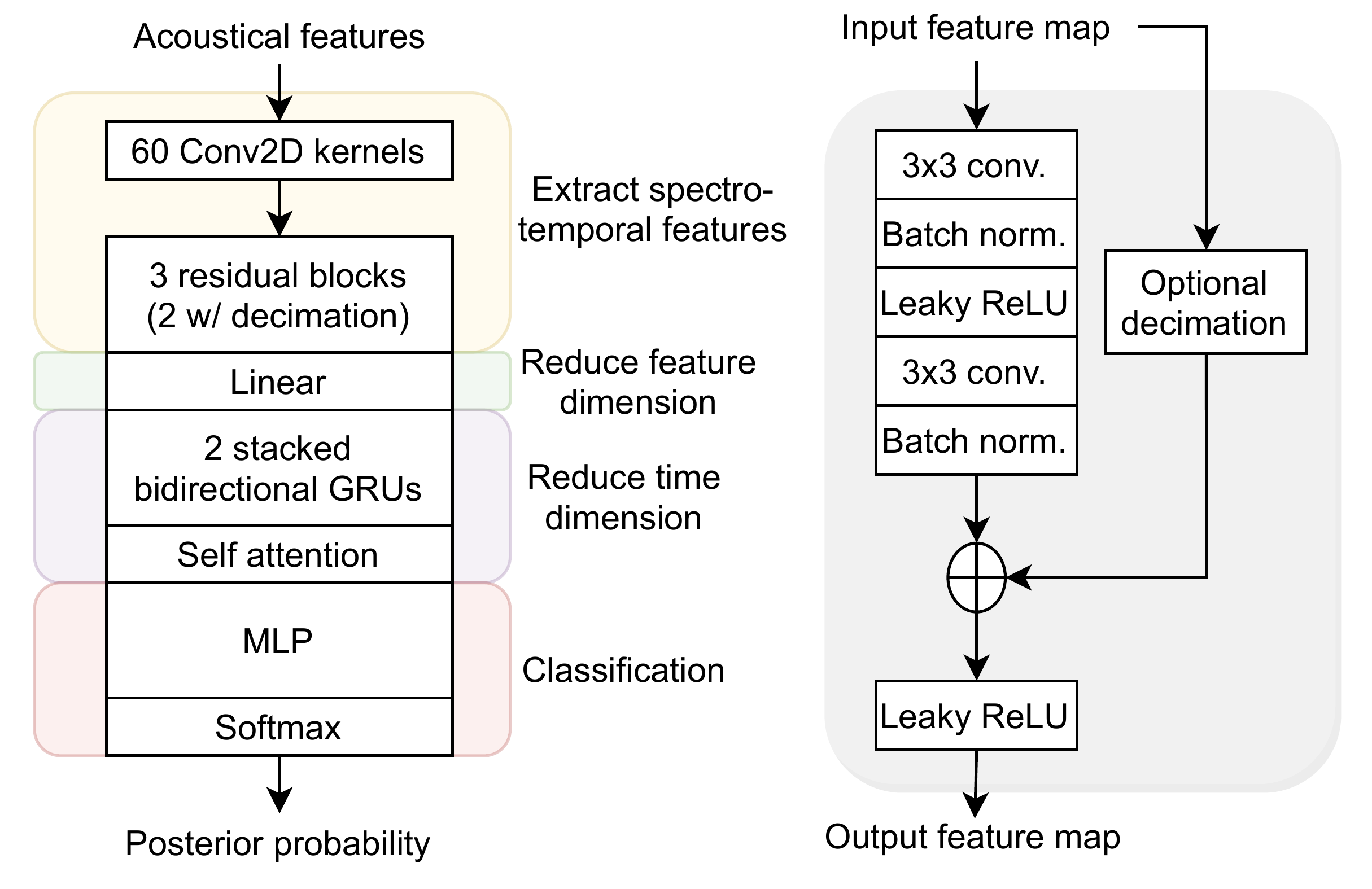}
  \caption{Block diagram of the baseline system (left) and the
zoomed-in view of one residual block (right).}
  \label{fig:baseline model}
  \vspace{-0.7cm}
\end{figure}

The baseline we use is a CNN-based model, similar to the baseline CNN model in \cite{vuong2020learnableVT}. As shown in Figure \ref{fig:baseline model}, the baseline model first consists of an initial convolutional layer followed by three residual blocks.  Next, the output is passed through bidirectional Gated Recurrent Units (GRUs) and a self-attentive pooling layer. After temporal modeling and the self-attentive pooling, the feature vector is passed through a one-hidden-layer multi-layer perceptron (MLP) with two dimensions for the output. Finally, softmax is applied to obtain the prediction probability of genuine speech. 


\subsection{Proposed feature}

The proposed feature is a simple and effective spectro-temporal feature: the 2D-DCT on log-Mel spectrogram. This is actually similar to the computation of Mel-frequency cepstral coefficients (MFCC) with the difference that we are applying a 2-dimensional (2D) discrete cosine transform (DCT) globally on both the temporal dimension and frequency dimension of the log-Mel spectrogram. The detailed computation steps are described as following: 
\begin{enumerate}
\item[a)] Employ the fast Fourier transform (FFT) to compute the spectrum $X(w)$ of $x(n)$.
\item[b)] Compute power spectrum $|X(w)|^2$ and obtain the Mel-spectrum $M$ by applying a Mel-frequency filter bank. 
\item[c)] Apply multi-dimensional discrete cosine transform (DCT) to log-Mel to obtain $dctn_M$. 
\item[d)] (Optional) Apply $l1$-normalization or standardization normalization on the obtained $dctn_M$. 
\end{enumerate}

\begin{figure}[t]
  \centering
  \includegraphics[width=0.88\linewidth]{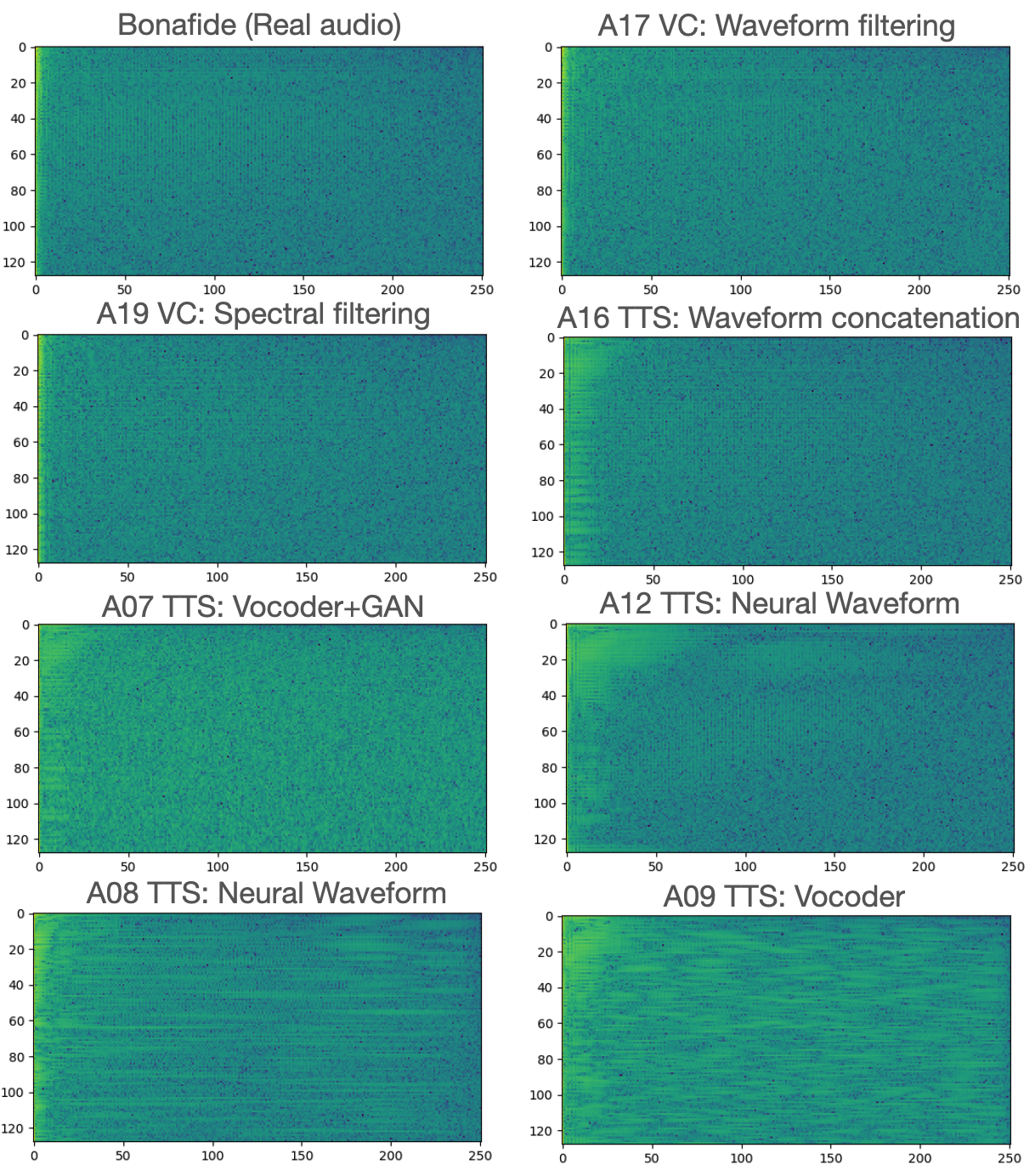}
  \vspace{-0.2cm}
  \caption{Visualization of the proposed features averaged within different spoofing types. Vertical axis is from mel-filters domain as spectro-modulation axis, and horizontal axis is from time frames as temporal-modulation axis. (Best viewed zoomed in)}
  \label{fig:vis}
  \vspace{-0.4cm}
\end{figure}

Figure \ref{fig:vis} shows the proposed 2D-DCT features for different spoofing types. The 2D-DCT features are in log-scale. From the visualization, we can see the proposed feature could obtain the differences in their patterns across different spoofing types. A17 and A19 use signal processing methods to generate fake audios, and the proposed features of these two are similar to the bonafide. In contrast, other methods give more complex changes compared to the bonafide (real audio) type.

\subsection{Implementation details}

For experiments with conventional and proposed features, we verify the spoofing countermeasures in performance improvements. We use the detection model that is modified from the residual net architectures proposed in \cite{alzantot2019deepASVspoofpipeline}. To evaluate the proposed features, this model is similar to our baseline model without the attention layer since the temporal information is already condensed into the global DCT domain. The audio sequences are cut or padded to 4 seconds, as the temporal range. The sampling rate is 16k, the FFT size is 1024, the window size is 512 and the hop size is 256, and the mel-filter number is 128. The details of the model implementation are in section 3.2 of \cite{gao2020detection}.   

Furthermore, we found the spectrum augmentation on the input features, and the normalization of the 2D-DCT features could improve the performance significantly, as shown in Table \ref{tab:tricks}. We implemented the SpecAugment (SA) \cite{park2019specaugment} approach on log-Mel spectrogram with torchaudio. The randomly masking on the frequency channels and time steps of the spectrogram helps preventing overfitting and increases the model's performance \cite{park2019specaugment}. For the SA on the proposed global modulation feature, a randomly zeroing-out manner is implemented to generate blank areas on both dimensions. This augmentation is only applied to the training data on the fly during training. For normalization on the 2D-DCT is applied using two approaches for comparison. Two normalization approaches, the l1-norm normalization and the mean/std standardization normalization, implemented using sklearn toolbox in PYTHON, achieve similar results. 
In contrast, the normalization does not help (much) for the other traditional features since the values are already in reasonable ranges and the l1-norm will break the spectral and temporal dynamics across the frames.  



\begin{table}[t]
  \caption{SpecAugment (SA) and Normalization approaches}
  \vspace{-0.35cm}
  \label{tab:tricks}
  \centering
  \begin{tabular}{lll}
    \toprule
    \textbf{Features}         &\textbf{t-DCF} &\textbf{EER (\%)}  \\
    \midrule
        log-Mel (Baseline$_1$) & 0.0902    &    6.551 \\
        \textbf{log-Mel   w/ SA} (Baseline$_2$)  & \textbf{0.0849}  &  \textbf{5.139}    \\
    \midrule    
        2D-DCT of log-Mel (Global M)        &   0.2851    &    12.40            \\
        Normalized Global M      & 0.1358   &    6.852        \\
        \textbf{Normalized Global M w/ SA (Ours)}   &    \textbf{0.1387}  &   \textbf{6.325}          \\
    \midrule
        T32 (Best single system \cite{todisco2019asvspoof}) & \text{0.1239} & 4.92 \\
    \bottomrule
  \end{tabular}
  \vspace{-0.75cm}
\end{table}

\begin{table}[t]
\caption{Single systems comparisons as ASV countermeasures}
\vspace{-0.7cm}
\label{tab: singlep}
\vskip 0.5 \baselineskip
\begin{center}
\scalebox{0.77}{ \begin{tabular}{l | cc | cc}
\toprule
\multicolumn{1}{c}{} & \multicolumn{2}{c}{Countermeasure EER\%} & \multicolumn{2}{c}{t-DCF} \\

\hline

\multicolumn{1}{c}{Features} & \multicolumn{1}{c}{DEV} & \multicolumn{1}{c}{EVAL} & \multicolumn{1}{c}{DEV} & \multicolumn{1}{c}{EVAL} \\
\hline
\midrule

\begin{tabular}{@{}c@{}@{}c@{}c@{}} Aperiodic parameters (AP) \\ Spectral envelope (SP) \\MFCC \\ CQCC \\ log-Mel spectrogram \\ Normalized Global M \\ Normalized Global M w/ SA \\ \end{tabular} 

& \begin{tabular}{@{}c@{}} 21.19 \\ 10.55\\ 7.14\\ 1.37 \\ 0.48 \\ 0.23 \\  0.17  \end{tabular} 

& \begin{tabular}{@{}c@{}} 20.65 \\ 9.31 \\ 11.64\\ 10.89 \\  9.39 \\ 6.85  \\ 6.32  \end{tabular} 

& \begin{tabular}{@{}c@{}} 0.4374 \\ 0.3520 \\ 0.1942 \\ 0.0407 \\ 0.0132  \\ 0.0067 \\ 0.0043   \end{tabular} 

& \begin{tabular}{@{}c@{}} 0.4445 \\ 0.2453 \\ 0.2663 \\ 0.2746 \\ 0.1954  \\ 0.1358 \\ 0.1387   \end{tabular} \\

\bottomrule 
\hline
\end{tabular}
}

\end{center}
\vspace{-0.5cm}
\end{table}

\section{Results}

\subsection{Single systems and weighted voting scores}




We evaluated the single system model taking in one type of feature and compared the proposed global modulation feature with the previously proposed feature aperiodic signal (AP), spectral envelope (SP) \cite{gao2020detection}, and other conventional features such as MFCC, CQCC, and log-Mel spectrogram. 
To have a fair comparison, the model is the same ResNet model as in Section 3.2 of \cite{gao2020detection} with the last layer's dimension change to facilitate the feature size difference. From the results in Table \ref{tab: singlep}, we can see the proposed feature is significantly better in both the EER and the t-DCF scores than the other features.

\begin{table}[t]
\caption{Weighted voting scores with different voting mechanisms}
\vspace{-0.6cm}
\label{tab: joint}
\vskip 0.5 \baselineskip
\begin{center}
\scalebox{0.8}{ \begin{tabular}{l | cc | cc}
\toprule
\multicolumn{1}{c}{} & \multicolumn{2}{c}{Global Modulation + Baseline$_1$} & \multicolumn{2}{c}{Global Modulation + Baseline$_2$ } \\

\hline

\multicolumn{1}{c}{Ratios} & \multicolumn{1}{c}{t-DCF} & \multicolumn{1}{c}{EER} & \multicolumn{1}{c}{t-DCF} & \multicolumn{1}{c}{EER} \\
\hline
\midrule

\begin{tabular}{@{}c@{}@{}c@{}c@{}} \textbf{min} \\ 0.0 \\ 0.1 \\ 0.2 \\ 0.3 \\ 0.4 \\ 0.5 \\ 0.6 \\ 0.7 \\0.8 \\0.9 \\ 1.0 \\ \textbf{max} \end{tabular} 

& \begin{tabular}{@{}c@{}} 0.1306 \\ 0.1397 \\ 0.1207 \\ 0.1063 \\ 0.0984 \\ 0.0923  \\ 0.0883 \\0.0867 \\ \textbf{0.0865} \\ 0.0870 \\ 0.0875 \\ 0.0902 \\ \textbf{0.0737} \end{tabular} 

& \begin{tabular}{@{}c@{}} 7.098 \\ 6.325 \\ 5.92 \\ \textbf{5.89} \\ 5.90 \\ 5.98 \\ 6.07 \\ 6.17 \\ 6.27 \\ 6.35 \\ 6.45 \\ 6.55 \\ \textbf{4.03} \end{tabular} 

& \begin{tabular}{@{}c@{}} 0.1230 \\ 0.1387 \\ 0.1253 \\  0.1141 \\ 0.1057 \\ 0.0994 \\ 0.0930 \\ \textbf{0.0890}\\0.1057 \\ 0.1142 \\ 0.1253 \\0.0849 \\ \textbf{0.0864} \end{tabular} 

& \begin{tabular}{@{}c@{}} 6.636 \\ 6.325 \\ 5.778 \\ 5.780 \\ 5.631 \\ 5.520 \\ 5.542 \\ \textbf{5.301} \\ 5.563 \\ 5.778 \\ 5.929 \\ 5.139 \\ \textbf{4.216} \end{tabular} \\

\bottomrule 
\hline
\end{tabular}
}

\end{center}
\vspace{-0.4cm}
\end{table}


We further evaluate the joint performance of our proposed feature with the strong baseline models. We use different voting mechanisms for the joint scores between the Global Modulation feature and the baseline models as following: For the prediction probability outputs of both systems, we weighted the prediction score using a ratio of 0.1 to 0.9. We use a max metric to keep the most confidence voting among the two systems, which gives us the best performance. In contrast, the min-metric keeps the lower confidence prediction of the two joint systems. From the results in Table \ref{tab: joint}, we can see the joint scores improve the overall countermeasure performance. 

\subsection{Audio type analysis}


To evaluate the detection performance on different spoofing audio types, we do a comprehensive analysis on the t-DCF and EER scores for all spoofing audio types in the LA evaluation set, as shown in Table \ref{tab: ID types}. The A17 type, generated with waveform filtering manipulations on the real audios, is visualized in Figure \ref{fig:vis}. It has a very similar modulation pattern to the bonafide audios and is the hardest type according to \cite{nautsch2021asvspoof}. Our baselines and the proposed feature achieve top performance, compared to the EERs of single systems reported in \cite{nautsch2021asvspoof}. And our joint system achieves one of the best three compared to all the other systems that use an ensemble of classifiers \cite{todisco2019asvspoof}.     

\begin{table*}
\caption{EERs of evaluation set for ASVspoof2019 LA for speaker verification}
\vspace{-0.5cm}
\label{tab:asv-eer}
\vskip 0.5 \baselineskip
\begin{center}
\scalebox{0.9}
{
\begin{tabular}{lc | cccccccccccccccc}
\toprule
\multicolumn{2}{c}{} & \multicolumn{13}{c}{ASV EER\%}\\
\hline
\multicolumn{2}{c}{Spoofing ID} & \multicolumn{1}{c}{A07} & \multicolumn{1}{c}{A08} & \multicolumn{1}{c}{A09} & \multicolumn{1}{c}{A10}  & \multicolumn{1}{c}{A11} & \multicolumn{1}{c}{A12} & \multicolumn{1}{c}{A13} & \multicolumn{1}{c}{A14} & \multicolumn{1}{c}{A15} & \multicolumn{1}{c}{A16} & \multicolumn{1}{c}{A17} & \multicolumn{1}{c}{A18} &
\multicolumn{1}{c}{A19} &
\multicolumn{1}{c}{ALL}\\
\hline
\midrule
& \begin{tabular}{@{}c@{}@{}c@{}c@{}} STFT\\ MFCC \\AP\\SP\\ \textbf{Global M}
\end{tabular} 
& \begin{tabular}{@{}c@{}} 2.33\\7.12\\38.93\\50.97\\1.45 \end{tabular} 
& \begin{tabular}{@{}c@{}} 2.65\\5.08\\32.46\\49.94\\ 8.01\end{tabular} 
& \begin{tabular}{@{}c@{}} 3.75\\8.12\\32.59\\40.07\\8.35\end{tabular} 
& \begin{tabular}{@{}c@{}} 47.56\\39.76\\42.37\\49.75\\31.97\end{tabular} 
& \begin{tabular}{@{}c@{}} 40.89\\28.99\\38.29\\49.25\\32.85\end{tabular} 
& \begin{tabular}{@{}c@{}} 47.59\\49.01\\43.28\\52.04\\38.92\end{tabular} 
& \begin{tabular}{@{}c@{}} 37.01\\33.81\\37.02\\52.30\\20.64\end{tabular} 
& \begin{tabular}{@{}c@{}} 29.09\\19.04\\33.96\\51.03\\14.10 \end{tabular} 
& \begin{tabular}{@{}c@{}} 35.48\\41.39\\41.12\\51.74\\28.22\end{tabular} 
& \begin{tabular}{@{}c@{}} 4.09\\9.08\\49.06\\51.99\\2.91\end{tabular} 
& \begin{tabular}{@{}c@{}} 12.07\\18.00\\40.05\\41.49\\23.93\end{tabular} 
& \begin{tabular}{@{}c@{}} 28.61\\16.47\\34.57\\46.16\\27.79\end{tabular} 
& \begin{tabular}{@{}c@{}} 1.88\\2.09\\44.53\\45.78\\1.11\end{tabular}  
& \begin{tabular}{@{}c@{}} 22.24\\15.99\\39.25\\42.08\\18.69\end{tabular} \\
\bottomrule 
\hline
\end{tabular}
 

}
\end{center}
\end{table*}
\begin{table*}
\caption{Breakdown analysis of the performance on different Spoofing audio types}
\vspace{-0.6cm}
\label{tab: ID types}
\vskip 0.5 \baselineskip
\begin{center}
\scalebox{0.8}{ \begin{tabular}{l | cc | cc | cc | cc | cc | cc }
\toprule
\multicolumn{1}{c}{} & \multicolumn{2}{c}{Info} & \multicolumn{2}{c}{Baseline$_1$} & \multicolumn{2}{c}{Baseline$_2$} & \multicolumn{2}{c}{Proposed feature} & \multicolumn{2}{c}{Joint w/ Baseline$_1$} & \multicolumn{2}{c}{Joint w/ Baseline$_2$} \\

\hline

\multicolumn{1}{c}{ID} & 
\multicolumn{1}{c}{System} & \multicolumn{1}{c}{Details} &
\multicolumn{1}{c}{t-DCF} & \multicolumn{1}{c}{EER} & \multicolumn{1}{c}{t-DCF} & \multicolumn{1}{c}{EER} & \multicolumn{1}{c}{t-DCF} & \multicolumn{1}{c}{EER} &
\multicolumn{1}{c}{t-DCF} & \multicolumn{1}{c}{EER} &
\multicolumn{1}{c}{t-DCF} & \multicolumn{1}{c}{EER} \\

\hline
\midrule

\begin{tabular}{@{}c@{}@{}c@{}c@{}} A07 \\ A08 \\ A09 \\ A10 \\ A11 \\ A12 \\ A13 \\ A14 \\ A15 \\ A16 \\ A17 \\ A18 \\ A19 \end{tabular} 

& \begin{tabular}{@{}c@{}} TTS \\ TTS \\ TTS \\ TTS \\ TTS \\ TTS \\ TTS-VC \\ TTS-VC \\ TTS-VC \\ TTS \\ VC \\ VC \\ VC \end{tabular} 

& \begin{tabular}{@{}c@{}} Vocoder+GAN \\ Neural waveform \\ Vocoder \\ Neural waveform \\ Griffin lim \\ Neural waveform \\ WC + waveform filtering \\ Vocoder \\ Neural waveform \\ Waveform concatenation (WC) \\ Waveform filtering \\ Vocoder \\  Spectral filtering \end{tabular} 

& \begin{tabular}{@{}c@{}} 0.0000 \\ 0.0463 \\ 0.0015 \\ 0.0084 \\ 0.0102 \\ 0.0041 \\ 0.0029 \\ 0.0079 \\ 0.0186 \\ 0.0007 \\ 0.9760 \\ 0.0061 \\ 0.0040  \end{tabular} 

& \begin{tabular}{@{}c@{}} 0.0000 \\ 1.4901 \\ 0.0577 \\ 0.3022 \\ 0.3667 \\ 0.1222 \\ 0.0985 \\ 0.2445 \\ 0.5942 \\ 0.0407 \\ 44.486 \\ 0.2037 \\ 0.1222 \end{tabular}

& \begin{tabular}{@{}c@{}} 0.0000 \\ 0.0163 \\ 0.0003 \\ 0.0058 \\ 0.0072 \\ 0.0020 \\ 0.0003 \\ 0.0037 \\ 0.0061 \\ 0.0005 \\ 0.7670 \\ 0.0098 \\ 0.0051  \end{tabular} 

& \begin{tabular}{@{}c@{}} 0.0000 \\ 0.5297 \\ 0.0170 \\ 0.2445 \\ 0.2852 \\ 0.0645 \\ 0.0170 \\ 0.1222 \\ 0.1799 \\ 0.0169 \\ 26.538 \\ 0.3259 \\ 0.1630 \end{tabular}

& \begin{tabular}{@{}c@{}} 0.0054 \\ 0.0521 \\ 0.0093 \\ 0.0417 \\ 0.0407 \\ 0.0635 \\ 0.0650 \\ 0.0270 \\ 0.0248 \\ 0.0062 \\ 0.9017 \\ 0.1985 \\ 0.0151  \end{tabular} 

& \begin{tabular}{@{}c@{}} 0.1799 \\ 1.9727 \\ 0.2852 \\ 1.3208 \\ 1.3038 \\ 1.9557 \\ 2.0372 \\ 0.8149 \\ 0.7911 \\ 0.1867 \\ 36.286 \\ 6.1286 \\ 0.5297 \end{tabular}

& \begin{tabular}{@{}c@{}} 0.0014 \\ 0.0147 \\ 0.0028 \\ 0.0080 \\ 0.0083 \\  0.0090 \\ 0.0113 \\ 0.0069 \\ 0.0069 \\ 0.0010 \\ 0.8004 \\ 0.0201 \\ 0.0050 \end{tabular} 

& \begin{tabular}{@{}c@{}} 0.0407 \\ 0.5297 \\ 0.0815 \\ 0.2852 \\ 0.2682 \\ 0.2852 \\ 0.3429 \\ 0.2274 \\ 0.2275 \\ 0.0407 \\ 28.324 \\ 0.6111 \\ 0.1799  \end{tabular}

& \begin{tabular}{@{}c@{}} 0.0020 \\ 0.0254 \\ 0.0035 \\ 0.0164 \\ 0.0152 \\ 0.0193 \\ 0.0218 \\ 0.0095 \\ 0.0097 \\ 0.0016 \\ 0.6218 \\ 0.0602 \\  0.0058 \end{tabular} 

& \begin{tabular}{@{}c@{}} 0.0645 \\ 0.7911 \\ 0.1392 \\ 0.5059 \\ 0.4720 \\ 0.6111 \\ 0.6689 \\ 0.3022 \\ 0.3259 \\ 0.0578 \\ 28.405 \\ 1.7927 \\ 0.2037 \end{tabular}\\

\bottomrule 
\hline
\end{tabular}
}

\end{center}
\vspace{-0.4cm}
\end{table*}

\subsection{Speaker verification using the proposed features}
To evaluate our proposed feature's effectiveness, we evaluate the feature under the automatic speaker verification scenario, as in~\cite{gao2020detection}. The ASV model is trained with the ASVspoof 2019 data LA training set. 
We assign each spoofed utterance an identity that uniquely incorporates both speaker and attack. The 20 speakers and 6 types of attack in the ASVspoof2019 LA training set are combined into 120 "spoofed identities". With the bonafide audios, we have positive pairs, and negative pairs generated randomly in a balanced 1:1 ratio. 
The results are shown in Table \ref{tab:asv-eer}. The proposed feature is compared with other features' results of \cite{gao2020detection}. Unlike AP and SP, the proposed 2D modulation feature is not only more powerful as in a detection model but also effective in the audio type and speaker verification tasks. This clearly shows the potential of this proposed feature for several applications.

\section{Discussions}

As the above results show, our proposed global modulation feature has a strong performance compared to other conventional features. We also test our best model's detection accuracy on the other external datasets FoR \cite{reimao2019FoRdataset} and RTVCspoof collected in \cite{subramani2020AIFlearning}. For each dataset, 200 fake and 200 real samples are selected randomly from their test sets. Our Global modulation feature model could also predict the class of the randomly selected test data with reasonable accuracy of 90\% to 98\%. 

We also compared the global modulation feature on the high-frequency section of the log-Mel spectrogram compared with the low-frequency section. Consistent with \cite{tian2016spoofinghl}, the high-frequency section gives higher detection performance compared to the low-frequency section, although still not as good as using the global information altogether. 
Finally, we compare a blocked version of the modulation feature with our proposed global modulation feature. We did a simple 2x2 division on the log-Mel spectrogram and computed the 2D-DCT features separately for each block. The resulting localized modulation features give a significantly lower detection performance of around 20\% EER. This shows the importance of taking long-range frequency computation to obtain the global inconsistencies for the audio detection leanings. Interestingly, in \cite{vuong2020learnableVT}, their proposed spectro-temporal receptive fields (STRFs) is a localized modulation feature. And in their experiments for the ASVspoof challenge, they concluded that 'the
STRFs effectively reject distractor noise, but are by themselves
not sufficient for discriminating real from synthetic speech'. Their results, in comparison, give another evidence for the importance of computing the modulation features globally.  

Also, it needs to be noted that the eval results for each feature are averaged across the eval EERs and t-DCFs of multiple runnings' best validation models for the soundness of the scores. The best eval score we have from a single running may be lower (E.g. the best baseline we have has an EER of 4.03\%). The t-DCF score is evaluated using the same metric as in \cite{todisco2019asvspoof}.

\section{Conclusions}

In this paper, we propose a simple yet effective feature, the global modulation feature, inspired by the fake audios' artifacts. We show that this proposed feature could improve the strongest baseline we have to further increase the countermeasure system's detection performance for the ASV system. Furthermore, we use this proposed feature to train our own ASV system and show that it also works very well for speaker verification tasks. This shows the broader potentials of the proposed global modulation feature.

In future works, we could embrace more data augmentation approaches, e.g., adding noise, etc. Moreover, with the future-released evaluation plan from ASVspoof challenge 2021, we would also evaluate the proposed feature's robustness to channel variations and its performance with the physical access (PA) dataset in ASVspoof Challenges \cite{wu2015asvspoof, wu2017asvspoof, todisco2019asvspoof, nautsch2021asvspoof}. 


\bibliographystyle{IEEEtran}
\bibliography{mybib}


\end{document}